# Emergent phenomena at oxide interfaces studied with standing-wave photoelectron spectroscopy

C.-T. Kuo,[1] G. Conti,[2,3] J. E. Rault,[4] C. M. Schneider,[3,5] S. Nemšák,[2,*] and A. X. Gray[6,*]

[1]Stanford Synchrotron Radiation Lightsource, SLAC National Accelerator Laboratory, Menlo Park, California 94025, USA
[2]Advanced Light Source, Lawrence Berkeley National Laboratory, Berkeley, California 94720, USA
[3]Department of Physics, University of California, Davis, Davis, California 95616, USA
[4]Synchrotron-SOLEIL, Saint-Aubin, BP48, F91192 Gif sur Yvette Cedex, France
[5]Peter Grünberg Institut (PGI-6), Forschungszentrum Jülich GmbH, D-52425 Jülich, Germany
[6]Department of Physics, Temple University, Philadelphia, Pennsylvania 19122, USA
[*]Corresponding authors: snemsak@lbl.gov, axgray@temple.edu

**ABSTRACT**

Emergent phenomena at complex-oxide interfaces have become a vibrant field of study in the past two decades due to the rich physics and a wide range of possibilities for creating new states of matter and novel functionalities for potential devices. Electronic-structural characterization of such phenomena presents a unique challenge due to the lack of direct yet non-destructive techniques for probing buried layers and interfaces with the required Ångstrom-level resolution, as well as element and orbital specificity. In this review article we survey several recent studies wherein soft x-ray standing-wave photoelectron spectroscopy, a relatively newly developed technique, is used to investigate buried oxide interfaces exhibiting emergent phenomena such as metal-insulator transition, interfacial ferromagnetism, and two-dimensional electron gas. Advantages, challenges, and future applications of this methodology are also discussed.

## I. INTRODUCTION

The x-ray standing-wave technique was pioneered by Boris Batterman in the early 1960s [1] and over the next several decades has been used extensively to tackle a wide variety of surface science problems, such as solving structures of adsorbates [2], investigating the localization of impurities [3], and surface reconstructions [4]. All of the above-mentioned studies utilized standing-wave formation by Bragg reflection from atomic planes and were carried out in the hard x-ray regime, which is ideally suited for atomic-resolution crystallographic studies due to the characteristic sub-nanometer wavelengths. The basic principles and history of the standing-wave technique, including notable seminal studies, are described in detail in the book by Zegenhagen and Kazimirov [5] as well as several recent topical reviews [6-8].

Standing-wave studies of the (buried) complex-oxide interfaces were, to a large degree, made possible by the extension of the standing-wave technique into the soft x-ray regime applied to artificial superlattices, which benefited greatly from the higher photoionization cross-sections, better energy resolution, and significantly easier collection of electron-momentum resolved data. Although the earliest studies by Yang and Fadley [9] and by Kim and Kortright [10] focused on magnetic interfaces, such as Fe/Cr and Co/Pd, and took advantage of the x-ray magnetic circular dichroism (XMCD) effects at the transition-metal $L$ edges and/or $2p$ core-level photoemission peaks, in the 2010s, following the discoveries of emergent phenomena [11-13] the focus shifted to strongly correlated perovskite oxides.

Over the recent decade the research group of Charles S. Fadley, celebrated in this focus issue, have utilized soft x-ray standing-wave spectroscopy in conjunction with other experimental techniques to study a wide variety of complex-oxide interface systems exhibiting emergent electronic and magnetic phenomena [14-22]. Their experimental progress was met by the collaborators on the theory and computation side with the development of advanced x-ray optical codes for fitting and optimization of the standing-wave data [23,24], as well as modeling of the depth-resolved electronic structure via advanced Green's function methods, such as one-step theory of photoemission [25,26]. In the latter part of the decade, the second generation of researchers originating from the Fadley group and its collaborators have continued this rich scientific tradition with the most recent studies of interfacial charge reconstructions [27], structural distortions [28], and electrochemically driven surface transformations [29].

In this article we overview several key studies from this period, with the specific focus on the depth-resolved investigations of the emergent interfacial phenomena in oxide systems. Among such phenomena, we focus on the interfacial metal-insulator transition in $LaNiO_3/SrTiO_3$ [15], the occurrence of a two-dimensional ferromagnetism in $LaNiO_3/CaMnO_3$ [27], the formation of a two-dimensional electron gas in $GdTiO_3/SrTiO_3$ [18,19], and the built-in electrostatic potential in $LaCrO_3/SrTiO_3$ superlattices [21]. We point out the importance of x-ray optical calculations and modeling in such studies, as well as the complementary role of first-principles theory for the interpretation of the experimental data. We also stress the strong synergy between x-ray standing-wave spectroscopy and other advanced characterization techniques, such as scanning transmission electron microscopy (STEM) with electron energy loss spectroscopy (EELS), x-ray absorption spectroscopy (XAS), as well as other modalities of photoelectron spectroscopies, such as hard x-ray photoemission (HAXPES) and soft x-ray angle-resolved photoelectron spectroscopy (SX-ARPES).

## II. Interfacial metal-insulator transition in $LaNiO_3/SrTiO_3$

Fundamental understanding and control of metal-insulator transitions (MIT) in strongly correlated material systems is of direct relevance and applicability to future device design [30]. One of the potential building blocks of such devices, lanthanum nickelate $LaNiO_3$, is one of the best-known examples of a complex transition-metal oxide in which electronic properties can be manipulated via thickness, strain, and heteroengineering [31-33]. In bulk form, unlike any other rare-earth nickelate compound in the $RNiO_3$ series, $LaNiO_3$ remains metallic at all temperatures [34]. However, in the form of a thin film or as a constituent layer in a superlattice, it shows insulating behavior at room temperature below a critical thickness of a few unit cells [35,36]. Numerous prior studies have addressed this phenomenon in single $LaNiO_3$ layers via a combination of advanced spectroscopic techniques and state-of-the-art theoretical calculations [31,37,38]. A seminal standing-wave photoemission study by Kaiser *et al.* demonstrated that, in superlattices, this phenomenon is not homogeneously distributed over the thickness of the $LaNiO_3$ layer but is more pronounced near the interfaces [15]. A follow-up momentum-resolved standing-wave ARPES study by Eiteneer *et al.* confirmed this observation and expanded on the details of the electronic structure with additional measurements and corroborating theoretical calculations [19].

For these studies, high-quality epitaxial superlattice samples consisting of ten bilayers of 4 unit cells (u.c.) thick $LaNiO_3$ and 3 u.c. thick $SrTiO_3$ were grown on a (001) $(LaAlO_3)_{0.3}\times(Sr_2AlTaO_3)_{0.7}$ (LSAT) substrate via rf magnetron sputtering. Standard characterization techniques, such as x-ray diffraction (XRD) and high-resolution STEM were used to verify the coherent epitaxy as well as the sharpness of the interfaces, while the results of the electronic transport measurements confirmed the metallicity of the superlattice. The standing-wave photoemission experiments were carried out at beamline 7.0.1 of the

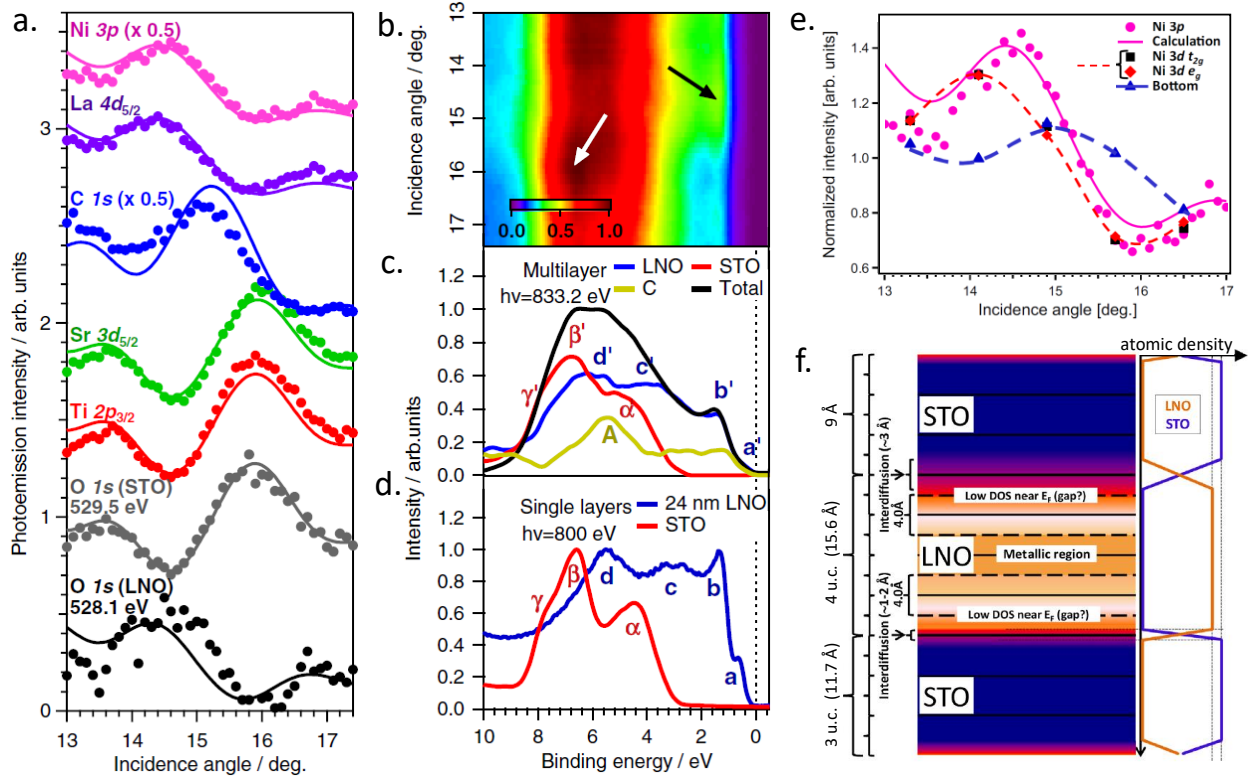

**FIG. 1.** a. Experimental standing-wave rocking curves (round markers) for all relevant core levels with best theoretical fits (solid curves). b. Intensity map of the valence band region for different incidence angles with the color scale corresponding to the photoemission yield. Two characteristic maxima split by 1.4° are marked by black and white arrows. c. Total valence-band spectrum (black curve) of the superlattice with deconvoluted spectral components showing the characteristic angular behavior of LaNiO$_3$ (blue curve), SrTiO$_3$ (red curve), and C-containing surface adsorbate (brown curve). d. Valence-band spectra of a SrTiO$_3$ single crystal (red curve) and a bulk 24 nm-thick LaNiO$_3$ film (blue curve). Characteristic features are labeled by lowercase Latin letters for LaNiO$_3$ and Greek letters for SrTiO$_3$. e. Integrated intensities of the Ni 3$d$ $t_{2g}$, $e_g$, and the band bottom, compared to the Ni 3$p$ core-level rocking curve (experiment and simulation). e. Schematic depth profile of the first few layers of the sample, as derived from the analysis of the core-level and the valence-band rocking curves. Reproduced with permission from [15] and [19].

Advanced Light Source and at the photon energy near the La 3$d_{5/2}$ absorption resonance (833.2 eV) in order to maximize the x-ray standing-wave node-to-antinode $E$-field contrast, as previously done by Gray *et al.* in the studies of manganite multilayers [14,17].

For the standing-wave measurements, the core-level intensities for each constituent element in the superlattice were measured and modeled as a function of the x-ray grazing incidence angle near the first-order Bragg condition as shown in Fig. 1a. The best fits (solid curves) to the experimental data (round symbols) obtained using an x-ray optical theoretical code [23] yielded an Ångstrom-level depth-resolved chemical profile of the sample, with some of the resultant values, such as layer thicknesses and interdiffusion lengths, shown in Fig. 1f.

The key result of the study by Kaiser *et al.* comes from the standing-wave measurement and spectral decomposition of the high-resolution valence-band spectra of the superlattice. Figure 1b shows the said valence-band spectra measured as a function of the x-ray grazing incidence angle and depicted as a two-dimensional color intensity map. In this representation, it is immediately evident that different spectral regions of the valence-band manifold exhibit markedly different angular behavior. For example, it is clear that the two main intensity maxima observed at the binding energies of approximately 1 eV and 7 eV and marked with the black and white arrows, respectively, occur at two angular positions that are separated by 1.4°, which roughly corresponds to the FWHM of the Bragg feature observed in the core-level SW rocking-curve measurements. This suggests that these two spectral components originate from different depths that are separated by one half-period of the superlattice. Additionally, a significant decrease in the spectral intensity is observed for angles greater than 16°, suggesting an opening of a bandgap in one of the layers.

It has been previously pointed out by Woicik *et al.* that the valence-band spectrum of a multilayer sample could be described to the first order of approximation by a simple superposition of the matrix-element-weighted projected densities of states (MEW-DOS) of the constituent layers [39]. Extrapolating from this observation, the measured valence-band intensities at each value of the binding energy should then show the angular dependence that could be described by a linear combination of the characteristic core-level rocking curves from the unique elements comprising the constituent layers. Thus, layer-specific valence-band spectra (MEW-DOS) could be deconvoluted from the valence-band spectrum of the superlattice sample by using the core-level SW rocking curves for Ti 2$p$ (representing SrTiO$_3$), La 4$d$ (representing LaNiO$_3$), and C 1$s$ (representing carbon-containing surface adsorbates) peaks.

The resultant layer-specific valence-band spectra for the LaNiO$_3$ and SrTiO$_3$ layers are shown in Figure 1c and compared to the corresponding valence-band spectra obtained separately from the single bulk-like films of LaNiO$_3$ and SrTiO$_3$ (Fig. 1d). While the SrTiO$_3$ spectra exhibit good agreement, a significant suppression of the near-Fermi-level Ni 3$d$ $e_g$ and $t_{2g}$ states (labeled a' and b') is observed for the LaNiO$_3$ layer in the superlattice sample, as compared to the bulk-like film. A further quantitative investigation of the angular behavior of the near-Fermi-level region of the spectrum and its comparison to the La 4$d$ (not shown here) and Ni 3$p$ rocking curves (see Fig. 1e) reveal that the spectral weights of the Ni 3$d$ $e_g$ and $t_{2g}$ states are preferentially suppressed within an approximately one-unit-cell thick region of the LaNiO$_3$ film near the interface with SrTiO$_3$. A schematic diagram of the calculated interface profile is shown in Figure 1f.

In summary, the study by Kaiser *et al.* [15] demonstrated a straight-forward methodology for deriving layer-resolved valence-band spectral functions in oxide superlattices, which has been since successfully applied to several other technologically relevant material systems [21]. Furthermore, together with a later work by Eiteneer *et al.*, [19] it has furthered our understanding of the interfacial MIT physics in Mott oxides.

### III.  Origins of interfacial ferromagnetism in LaNiO$_3$/CaMnO$_3$

From the early days of standing-wave photoelectron spectroscopy it has been suggested that the standing-wave excitation could be used in conjunction with such techniques as XMCD and STEM-EELS for the depth-resolved profiling of interfaces between magnetic materials [40]. Consequently, over the past twenty years, several such studies have been carried out on the heterostructures of key importance in the fields of spintronics and magnetic data storage. Examples of the investigated material systems include Fe/Cr [9], Fe/MgO [41], and CoFeB/CoFe [42] interfaces wherein the depth-dependence of electronic, chemical, and magnetic profiles have both scientific and technological significance.

With the advent of atomic-level oxide heteroengineering, standing-wave photoemission emerged as an ideal tool for studying new magnetic ground states at interfaces between strongly correlated oxides and the underlying electronic phenomena that are responsible for stabilizing these states. Due to its Ångstrom-level depth resolution, element and orbital specificity, and non-destructive nature, this technique is a formidable competitor of STEM-EELS and, when used in tandem with advanced microscopic techniques [17,28], provides a comprehensive picture of the structural and electronic interactions at the interface.

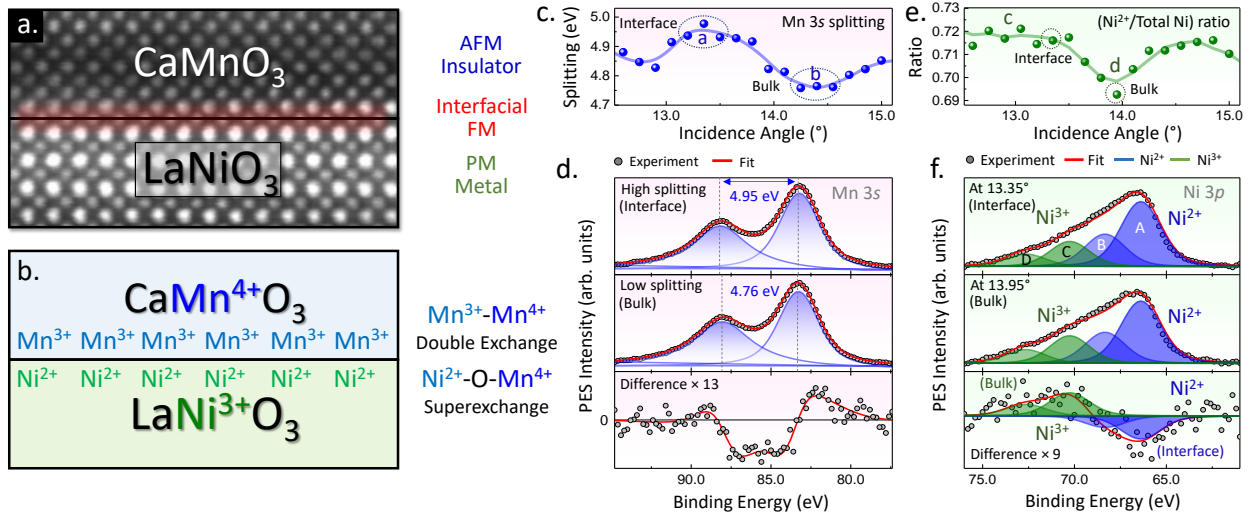

**FIG. 2.** a. High-resolution STEM-HAADF cross-sectional image of the coherent epitaxial $LaNiO_3/CaMnO_3$ interface. b. Schematic diagram of the proposed (and observed) interfacial charge reconstruction leading to the emergence of the electronic environment that is favorable for the stabilization of interfacial ferromagnetism. c. Depth dependent evolution of the Mn 3$s$ multiplet splitting as a function of the x-ray grazing incidence angle. d. Mn 3$s$ spectra recorded in the interface-sensitive and bulk-like-sensitive experimental geometries, showing an increase of the splitting, which is a spectroscopic signature of the reduced formal valency of Mn cations. e. Plot of the relative $Ni^{2+}$ peak(s) intensity as function of x-ray grazing incidence angle, with the interface-like and bulk-like Ni 3$p$ peaks shown in f. Reproduced with permission from [27].

Most recently, Chandrasena *et al.* [27] utilized soft x-ray standing-wave photoelectron spectroscopy in conjunction with STEM-EELS to demonstrate a depth-dependent charge reconstruction that leads to the emergence of interfacial ferromagnetism at the interface between paramagnetic $LaNiO_3$ and antiferromagnetic $CaMnO_3$ (Fig. 2a). The stabilization of a long-range ferromagnetic order at this interface system has been discovered nearly a decade ago [43] and was consequently attributed to two distinct mechanisms: a $Mn^{4+}$-$Mn^{3+}$ double exchange interaction in the interfacial $CaMnO_3$ layer and a $Ni^{2+}$-O-$Mn^{4+}$ superexchange interaction at the interface between $LaNiO_3$ and $CaMnO_3$ [44,45]. These studies also revealed presence of the necessary off-stoichiometric cations ($Ni^{2+}$ and $Mn^{3+}$) in the $LaNiO_3/CaMnO_3$ heterostructures using depth-averaging techniques such as XAS and XMCD. However, until recently, no direct measurement showed unambiguously that they were localized at the interface (as shown schematically in Fig. 2b), where the emergent magnetic state was observed via polarized neutron reflectivity [43].

In order to address the question of the underlying electronic origin of interfacial ferromagnetism in this material system, a high-quality epitaxial [4-u.c. $LaNiO_3$/4-u.c. $CaMnO_3$]×15 superlattice was synthesized on top of a single-crystalline $LaAlO_3$ (001) substrate using pulsed laser deposition. Coherent epitaxy, nominal thickness, and interface quality were verified using a combination of STEM, XRD, and soft x-ray reflectivity. Bulk magnetization measurements were used to verify presence of the ferromagnetic state, and depth-averaged XAS spectroscopy was used to confirm presence of the mixed valence states for both Mn (4+ and 3+) and Ni (3+ and 2+). After these preliminary characterization measurements, standing-wave photoelectron spectroscopy measurements were carried out at the high-resolution ADRESS beamline of the Swiss Light Source at the photon energy near the La $3d_{5/2}$ absorption resonance, as described in the previous section.

Core-level photoemission intensities from every constituent element in the superlattice were recorded as a function of grazing incidence angle near the first-order Bragg condition (approximately 14°) and self-consistently fitted using an x-ray optical theoretical code [23], yielding a detailed chemical profile of the sample, including thicknesses of all layers, interfacial interdiffusion lengths, as well as the angle-dependent profile of the standing-wave *E*-field intensity within the superlattice (see Figs. 2 and 3 in Ref. 27). Depth-resolved valence-state profiles of the Mn and Ni cations near the interface were extracted from the high-resolution spectra of the Mn 3*s* and Ni 3*p* peaks that were recorded as a function of the standing-wave position within the sample (Figs. 2c-f). Specifically, the depth-dependent evolution of the Mn 3*s* multiplet splitting (Fig. 2c) was utilized to determine the change in the Mn valence state [46] between the bulk-like and interface-like regions of the $CaMnO_3$ layer (Fig. 2d). Conversely, the relative peak intensities of the 3+ and 2+ components of the Ni 3*p* core level [47] were used to track the depth-dependent evolution of the formal valency of the Ni cations in the $LaNiO_3$ layer (Figs. 2e-f).

The background-subtracted and fitted data shown in Figures 2d,f revealed a depth-dependent charge redistribution within the $LaNiO_3/CaMnO_3$ heterostructure, with an increased concentration of the off-stoichiometric B-site cations ($Mn^{3+}$ and $Ni^{2+}$) near the interfaces and a higher abundance of "formal valence" cations ($Mn^{4+}$ and $Ni^{3+}$) in the bulk-like regions of the respective layers, as shown schematically in Fig. 2b. On the $CaMnO_3$ side of the interface, such charge reconstruction has been predicted to originate due to the leakage of itinerant 3*d* $e_g$ electrons from the adjacent Ni sites in $LaNiO_3$ [43,48]. Conversely, presence of the $Ni^{2+}$ cations on the $LaNiO_3$ side of the interface is consistent with the emergence of oxygen vacancies driven by the interfacial polar compensation mechanism [49]. This results in an electronic environment favorable for the stabilization of interfacial ferromagnetism within a narrow (one unit cell) region at the interface and mediated by the $Mn^{4+}$-$Mn^{3+}$ double exchange and/or $Ni^{2+}$-O-$Mn^{4+}$ superexchange interactions. The competition between these two distinct mechanisms can be influenced by varying the thickness of $LaNiO_3$, which undergoes a metal-insulator transition in the ultrathin (few-unit-cell) limit [44,45].

In summary, the study by Chandrasena *et al.* [27] established a powerful recipe for probing unit-cell-resolved changes in the interfacial valence states resulting in the stabilization of long-range ferromagnetic order at oxide interfaces. Future works will likely extend this methodology to the polarization- and momentum-dependent studies of such material systems, thus enabling rational strategies for designing functional Mott oxide heterostructures wherein magnetic states could be controlled via thickness, strain, ionic defects, and external stimuli.

### IV. Character of two-dimensional electron gas (2DEG) at the $GdTiO_3/SrTiO_3$ interface

One of the biggest strengths of standing-wave photoelectron spectroscopy is its ability to specifically probe electronic properties of buried interfaces, whether they are solid/solid or, with the help of special instrumentation, solid/liquid and solid/gas [50,51]. There has been a high interest in multilayer structures involving metal oxides, due to the novel electronic states that can develop at the interfaces between the different constituents, which are often very different from those states in the native bulk materials. A classic example of these phenomena is the two-dimensional electron gas (2DEG) at the $SrTiO_3/LaAlO_3$ interface [52], which was discovered to be superconducting [53], and also exhibits magnetic properties [54], in spite of both constituents being non-magnetic insulators. There are numerous studies that explore this system using ultraviolet and soft x-ray angle-resolved photoelectron spectroscopy as a prime tool to investigate electronic structure of solid materials [55], including a standing-wave photoelectron spectroscopy study exploring the role of oxygen vacancies [22].

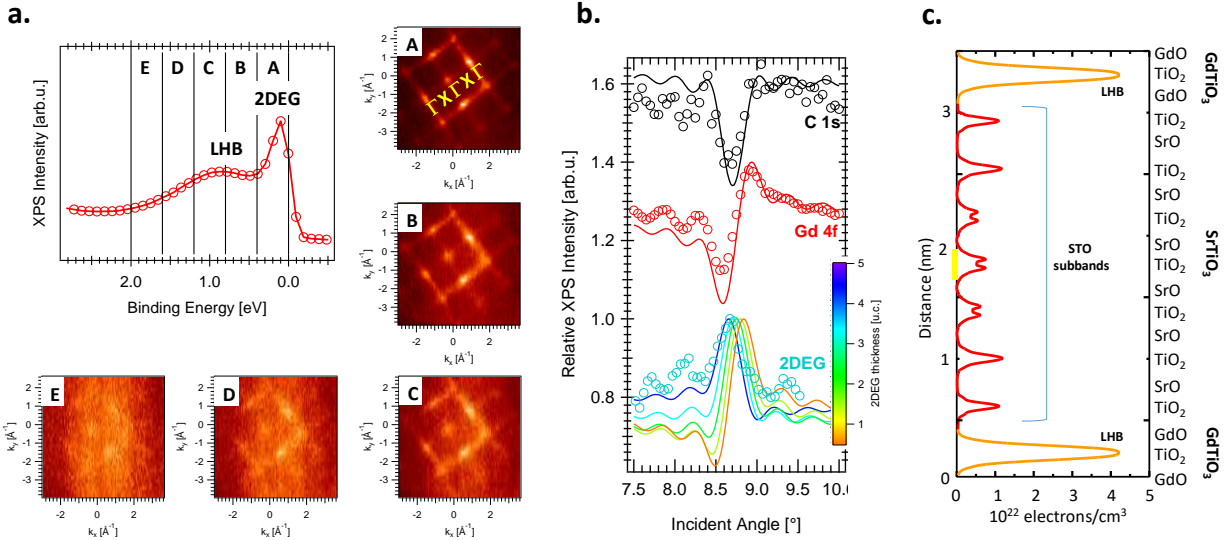

**FIG. 3.** a. Angle integrated (top left) and angle-resolved X-ray photoelectron spectroscopy data of GdTiO$_3$/SrTiO$_3$ superlattice valence band measured at resonant excitation energy of 465.2 eV. b. Experimental (open circles) and simulated (solid lines) rocking curves for contaminant species (C 1s), Gd 4f and valence band peak identified as 2DEG. 2DEG plot shows calculated rocking curves for different thicknesses of 2DEG ranging from 1 to 5 unit cells. c. Depth resolved sheet electron density as calculated by first-principle calculations, confirming the presence of 2DEG in the entire thickness of SrTiO$_3$ parts of the superlattice. Reproduced with permission from [18].

Other systems exhibiting interfacial 2DEG were discovered soon after. The GdTiO$_3$/SrTiO$_3$ heterostructure, for example, was studied extensively by Stemmer and collaborators, and its 2DEG exhibits extremely high carrier densities and ferromagnetic effects, with both electrostatic and doping modulation being observed to change the carrier properties [56,57]. In contrast to the case of SrTiO$_3$/LaAlO$_3$, the explanation for the emergence of 2DEG in GdTiO$_3$/SrTiO$_3$ can be found in a simple charge transfer picture [58]. Specifically, the GdO and TiO$_2$ planes in GdTiO$_3$ carry +1 and −1 formal ionic charges, respectively. Each GdO layer in GdTiO$_3$ can be considered to donate one-half an electron to the TiO$_2$ planes above and below it, including the interfacial TiO$_2$ plane, which is shared with SrTiO$_3$ and in which the half-electron forms a mobile 2DEG state. Experimentally, angle-resolved photoemission provided a detailed picture of the nature of the 2DEG (including exploration of critical thickness of SrTiO$_3$) [59], but it was an experiment with standing-wave excitation that revealed the thickness and location of 2DEG states relative to the interface [18].

In this study by Nemsak *et al.*, a superlattice of [6 u.c. SrTiO$_3$/3 u.c. GdTiO$_3$]×20 was prepared by molecular beam epitaxy and electronic transport measurements confirmed the presence of the interfacial 2DEG. The spectral and momentum signature of the 2DEG in GdTiO$_3$/SrTiO$_3$ is shown in Figure 3a. Two distinct states are resolved in the proximity of the Fermi edge. A sharp state right at the Fermi energy is identified as 2DEG, while a deeper broader peak is assigned to the Lower Hubbard Band (LHB) of the Mott-insulator GdTiO$_3$. The energy range of 0 to 2 eV binding energy is separated into five regions, labeled A-E, and the momentum-resolved dispersive features for each of the energy regions are shown. The 2DEG dispersion follows the symmetry of the crystal, with the electron density pockets connecting Γ points of neighboring Brillouin zones. Interestingly, the deeper-located bands corresponding to the LHB are showing a similar momentum dispersion, suggesting a somewhat similar character of these two states.

The exact depth location of the 2DEG is then studied by standing-wave photoemission. Several new approaches were used in this study, with one of them being the use of excitation energies just below and above a strong absorption edge (Gd $M_5$-edge in this case). Such choice of energies yields two effects – enhancement of the standing-wave modulation due to increased contrast (this was successfully demonstrated in $La_{0.7}Sr_{0.3}MnO_3$/$SrTiO_3$ study by Gray *et al.* earlier [14]) and the inversion of the standing-wave phase, practically shifting nodes and antinodes of the standing wave by almost half a period within the superlattice. Figure 3b shows experimental and calculated rocking curves for the excitation energy just below the $M_5$ edge of Gd (1181 eV). Surface contaminants (represented by C 1*s*), Gd 4*f*, and a spectral feature (state A, Figure 3a) corresponding to the 2DEG are shown.

The agreement between calculated and experimental curves for C 1*s* and Gd 4*f* confirms that the theoretical model used to simulate standing-wave data agrees very well with the experimental results, giving a credibility to the further analysis of the 2DEG location. The simulation of the 2DEG position revealed that it is not exclusively contained at the $GdTiO_3$/$SrTiO_3$ interface but expands throughout the entire thickness of the $SrTiO_3$ layers. Similar modulations (rocking curves) for the LHB states at 0.5 - 2 eV binding energy (not shown) and Gd 4*f* then suggest that LHB is located in $GdTiO_3$. These results are corroborated by first-principle calculations (shown in Figure 3c), confirming that states related to the LHB are present exclusively in $GdTiO_3$ and the 2DEG occupies $SrTiO_3$ in its entire thickness.

In summary, the study by Nemsak *et al.* [18] provided an invaluable information on the character of 2DEG, not only from the perspective of the electrons' energy and momentum, but also physical location of these electrons in the superlattice. The enhancement of the standing wave measurements by using excitation energies below and above the absorption edge provides another way of tailoring depth profile of the exciting x-rays and it has been since used in other studies [21]. Lastly, the combination of soft and hard x-ray spectroscopy together with first-principle calculations demonstrated the synergy of such instrumentally complex studies, which are able to answer fundamental and very complicated questions on the character of the emergent electronic states at buried interfaces.

## V. Direct measurement of the built-in electrostatic potential in $LaCrO_3$/$SrTiO_3$ superlattices

Epitaxial interfaces and superlattices of polar and nonpolar perovskite oxides have generated a considerable interest because they possess desirable properties for functional devices. For instance, Comes *et al.* demonstrated the existence of induced polarization fields throughout the $SrTiO_3$ layers of a $LaCrO_3$/$SrTiO_3$ superlattice by controlling the interfaces between polar $LaCrO_3$ and non-polar $SrTiO_3$ [60]. The induced polarization leads to built-in potential gradients within the $SrTiO_3$ and $LaCrO_3$ layers of the superlattice. Using standing-wave photoelectron spectroscopy, Lin *et al* explored the role of the two distinct and controllably charged interface structures [$(LaO)^+/(TiO_2)^0$ and $(SrO)^0/(CrO_2)^-$] due to a polar discontinuity along the [001] direction, and demonstrated that standing-wave photoelectron spectroscopy uniquely determines the built-in potential, along with the depth-dependent composition, and the electronic structure [21].

The studied superlattice consisted of [5 u.c. $LaCrO_3$/10 u.c. $SrTiO_3$]×10 and was prepared by oxide molecular beam epitaxy. Figures 4a and 4b show the experimental (open circles) and simulated (solid lines) rocking curves of the representative elemental states at photon energies of 829.7 and 831.5 eV, respectively. The high quality of the superlattice is confirmed by the core-level rocking curves, which exhibit almost identical intensity profiles for the atomic species that reside in the same layer (for example La 4*d* and Cr 3*p*, or Sr 3*d* and Ti 2*p*). On the other hand, the rocking curves corresponding to different layers, such as La 4$d$ and Sr 3$d$, are out of phase with respect to each other. From the theoretical simulations of the standing wave data, it can be concluded that the interfaces have a roughness of ~2–3 Å, and consist of alternating positively and negatively charged structures, specifically: (LaO)$^+$/(TiO$_2$)$^0$ with positive charge at the LaCrO$_{3(top)}$/SrTiO$_{3(bottom)}$ interface, and (SrO)$^0$/(CrO$_2$)$^-$ with negative charge at the SrTiO$_{3(top)}$/LaCrO$_{3(bottom)}$ interface. These results are also consistent with the STEM-EELS studies reported in a previous work by Comes *et al*. [61] The valence-band spectrum of the superlattice sample is the sum of the matrix-element-weighted densities of states (MEW-DOS) for all constituent layers with appropriate attenuation factors due to the electron scattering in the sample. Therefore, standing-wave modulations in the valence band exhibit a very complex behavior based on the character (and physical depth location) of the individual states. By using the core-level rocking curves as base vectors for projections of the valence band standing-wave data, one can distinguish the valence band contributions from the LaCrO$_3$ or SrTiO$_3$ components of the superlattice, as was earlier demonstrated on the case of LaNiO$_3$/SrTiO$_3$.

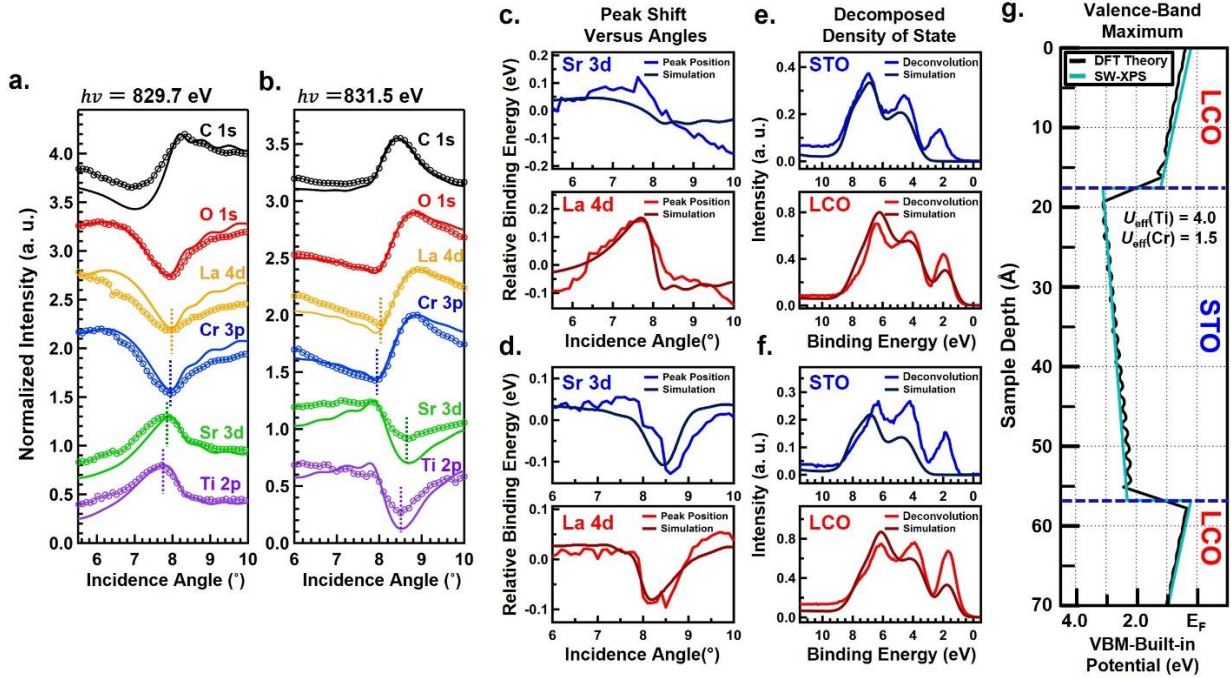

**FIG. 4.** Experimental (open circles) and simulated (solid) rocking curves of representative elemental states at the photon energies of a. 829.7 eV and b. 831.5 eV. The dashed vertical lines indicate the phase difference of the rocking curves. Experimental and simulated relative peak shifts for Sr 3$d$ and La 4$d$ core levels versus the incidence angle at the photon energies of c. 829.7 eV and d. 831.5 eV. Experimental valence-band decompositions, showing the contributions from the SrTiO$_3$ and LaCrO$_3$ layers, and corresponding simulations using XPS reference spectra from bulk SrTiO$_3$ and thick-film LaCrO$_3$, at the photon energies of e. 829.7 eV and f. 831.5 eV. g. SW-XPS-derived (turquoise curves) and DFT-calculated (PBEsol) depth-resolved valence-band maxima (black curves) for the top three layers of the LaCrO$_3$/SrTiO$_3$ superlattice. Reproduced with permission from [21].

For determining the built-in potential depth profile in the superlattice, the analysis procedures involve an investigation of the apparent core-level binding-energy shifts as the standing wave is scanned through the superlattice. Figures 4c and 4d show the experimental and simulated peak shifts for the Sr 3$d$ and La 4$d$ core levels versus incidence angle at photon energies of 829.7 and 831.5 eV. The experimental Sr 3$d$ and La 4$d$ peak positions change in binding energy on the order of 0.1–0.2 eV around the Bragg angle. Moreover, the form of these rocking curves is quite different for the two x-ray energies due to the difference of the standing-wave phases. In the simulated peak shifts, the depth integrated Sr 3*d* and La 4*d* spectra were calculated over the entire range of the angular scans and their angular dependences are labeled as "*Simulation*" in figure panels 4c and 4d. The simulation assumed that the core-level binding energy follows only the built-in potential (hence there are no chemical shifts involved) at each depth and this potential could be described as a linear variation within each layer. The solution of the potential gradients was determined by least-square fitting. The procedure yields an excellent agreement between experiment and theory, with the exception of Sr 3*d* measured at the excitation energy of 829.7 eV and showing less variation in theory. The potential gradients, labeled as "*SW-XPS*" (turquoise curves in LaCrO$_3$ or SrTiO$_3$ layers), are shown in Fig. 4g.

The energy steps or valence-band offsets at each interface are also shown in Fig. 4g (blue dashed lines in sample depth of ~18 and 57 Å) and they are further expanded by the analysis of the valence-band maxima. The deconvoluted experimental spectra of the SrTiO$_3$ layer and LaCrO$_3$ layers are reported in Figs. 4e and 4f. In these two panels, the curves denoted "*Simulation*" are based upon summing the depth-attenuated bulk-like XPS reference spectra over the entire depth of the sample shifted by the corresponding total potential (Fig. 4g). The total potential $E_b^0(z_i)$ includes the potential gradients within constituent layers and steps at the polar interfaces. By combining the derivation of the slopes of the electrostatic potential within each layer and the magnitudes of the valence band offsets, we finally determine the absolute potential value with respect to the valence-band maxima, annotated as the SW-XPS derived profile in Fig. 4g. The results of the analysis yield a clear agreement between the qualitative expectation of the charged-interface configuration and the signs of the potential gradients: higher (lower) binding energy for valence electrons at the positively (negatively) charged interfaces. The experimental results are corroborated by using density-functional theory (DFT) simulations with the PBEsol density functional [62], as implemented in the VASP code [63] with an adjustable U$_{eff}$ parameter for *d-d* correlation in both layers, and these results (black curves in Fig. 4g) agree in a remarkable way with the experimental results for the slopes and the offsets at the interfaces.

In summary, the study by Lin *et al.* [21] demonstrated the use of soft x-ray SW-XPS to extract the depth-resolved atomic and electronic structure, and the built-in potential in the LaCrO$_3$/SrTiO$_3$ superlattice. Furthermore, the analysis of apparent binding energy shifts of the core-levels and of the deconvoluted valence-band spectra, as compared to theoretical simulations, has permitted determination of the built-in potential variation with depth in great detail, including the band offsets at the polar interfaces. The experimental results are in excellent agreement with DFT, demonstrating the power and great synergy of using this experimental determination in tandem with first-principle calculations.

## VI. SUMMARY AND FUTURE OUTLOOK

In this review article we have highlighted and summarized several recent studies wherein soft x-ray standing-wave photoemission spectroscopy, as pioneered by Charles S. Fadley and collaborators, was utilized for the depth-resolved investigations of the emergent phenomena at the interfaces between strongly correlated perovskite oxides. Specifically, we discussed unique advantages of the valence-band SW-XPS and SW-ARPES for the studies of interfacial metal-insulator transitions, the electronic origins of interfacial ferromagnetism, two-dimensional electron gas, and built-in potential gradients in oxide superlattices. Strong synergy with complementary spectroscopic and microscopic techniques was emphasized, as well as the key roles of x-ray optical modeling and first-principles theoretical methods, such as DFT and one-step theory of photoemission.

We believe that the future holds many more exciting developments in the field of standing-wave photoemission, with an ever-widening variety of materials and interfaces that challenge us to understand and control the most fundamental physical processes, and at the same time point to potential solutions to some of our most formidable technological and societal problems. Adding to the array of new instruments at our disposal are the x-ray free-electron lasers (FEL), providing x-ray pulses that are Fourier-transform limited in the femtosecond regime. In particular, the new high-repetition rate FELs, such as LCLS-II and XFEL, will permit dealing more effectively with sample radiation damage, as well as space-charge effects in photoemission, leading to the addition of time resolution to standing-wave spectroscopy and thus further expanding the range of questions accessible to this powerful experimental technique.

**ACKNOWLEDGEMENTS**

The authors would like to express their deepest gratitude to the late Prof. Charles S. Fadley, a pioneer in the field of photoelectron spectroscopy, who inspired generations of scientists and educators worldwide. The specific sources of funding for the various studies presented here are listed in the publications cited. Beyond this, A.X.G. acknowledges support from the U.S. Department of Energy, Office of Science, Office of Basic Energy Sciences, Materials Sciences and Engineering Division under Award No. DE-SC0019297. The use of the Stanford Synchrotron Radiation Lightsource, SLAC National Accelerator Laboratory, was supported by the U.S. Department of Energy, Office of Science, Office of Basic Energy Sciences, under Contract No. DE-AC02-76SF00515. J.E.R. acknowledges the support of a public grant from the "Laboratoire d'Excellence Physics Atom Light Matter" (LabEx PALM) overseen by the ANR as part of the "Investissements d'Avenir" Program (Ref. No. ANR-10-LABX-0039). We acknowledge SOLEIL for provision of synchrotron radiation facilities.

**DATA AVAILABILITY**

Data sharing is not applicable to this article as no new data were created or analyzed in this study.